\documentclass[twocolumn,showpacs,amsmath,amssymb,prl,superscriptaddress]{revtex4}

\usepackage{graphicx}
\usepackage{dcolumn}
\usepackage{bm}
\usepackage{amssymb}
\usepackage{bbm}
\usepackage{calc}

\newcommand{\ii}{ {\rm i} }

\newcommand{\eq}[1]{Eq.~(\ref{#1})}

\begin{document}

\title{Coherent instabilities in a semiconductor laser with fast gain recovery}
\author{Christine Y. Wang}
\affiliation{Department of Physics, Harvard University, Cambridge,
Massachusetts 02138, USA}
\author{L. Diehl}
\affiliation{Division of Engineering and Applied Sciences, Harvard
University, Cambridge, Massachusetts 02138, USA}
\author{A. Gordon}
\author{C. Jirauschek}
\author{F. X. K\"artner}
\email{kaertner@mit.edu} \affiliation{Department of Electrical
Engineering and Computer Science and Research Laboratory of
Electronics, Massachusetts Institute of Technology, 77 Massachusetts
Ave. Cambridge, Massachusetts 02139}
\author{A. Belyanin}
\affiliation{Department of Physics, Texas A \& M University, College
Station, Texas 77843, USA}
\author{D. Bour}
\author{S. Corzine}
\author{G. H\"ofler}
\affiliation{Agilent Technologies, Palo Alto, California 94306, USA}
\author{M. Troccoli}
\affiliation{Division of Engineering and Applied Sciences, Harvard
University, Cambridge, Massachusetts 02138, USA}
\author{J. Faist}
 \affiliation{Institute of Physics, University of Neuch\^atel, CH-2000 Neuch\^atel,
 Switzerland}
 \author{Federico Capasso}
 \email{capasso@deas.harvard.edu}
 \affiliation{Division of Engineering and Applied Sciences, Harvard
University, Cambridge, Massachusetts 02138, USA}

\begin{abstract}
We report the observation of a coherent multimode instability in
quantum cascade lasers (QCLs), which is driven by the same
fundamental mechanism of Rabi oscillations as the elusive
Risken-Nummedal-Graham-Haken (RNGH) instability predicted 40 years
ago for ring lasers. The threshold of the observed instability is
significantly lower than in the original RNGH instability, which we
attribute to saturable-absorption nonlinearity in the laser.
Coherent effects, which cannot be reproduced by standard laser rate
equations, can play therefore a key role in the multimode dynamics
of QCLs, and in lasers with fast gain recovery in general.
\end{abstract}


\maketitle

 The fundamental coherent mechanism  that can destabilize a
single-mode laser was predicted in the early 60s \cite{Oraevsky} and
was later extended to multi-mode lasers \cite{Risken,Haken} where it
became known as the RNGH instability. These instabilities became
classic landmarks for the general field of nonlinear dynamics
\cite{haken3,khanin} because they emerge in conceptually the
simplest laser model, which in the single-mode case was shown to be
equivalent to the Lorentz model of deterministic chaos
 \cite{Haken2}. Another feature that makes these instabilities so interesting and unique
  is their coherent nature that involves the polarization of the medium as a nontrivial
dynamical variable. Most other physical mechanisms that can drive a
laser from a single-mode to a multi-mode regime, such as spatial and
spectral hole burning, Q-switching, and saturable absorption
\cite{Yariv, HausReview}, can be adequately described within the
standard rate equation formalism, in which the polarization of the
active medium is adiabatically eliminated. The RNGH instability and
its single-mode twin cannot be explained by the rate equations. Such
coherent effects can be only observed when the polarization is
driven  faster than or comparable to the dephasing time $T_2$
\cite{Eberly}.

 The origin of both phenomena is
the oscillation of the population inversion at the Rabi frequency
$\Omega_{\rm Rabi}$ that takes place when the intracavity laser
intensity becomes large. This results in a modification of the gain
spectrum and the emergence of sidebands separated from the maximum
of the gain curve by an amount corresponding to the Rabi frequency.
These sidebands can be regarded as a manifestation of parametric
gain. The instability sets in when the intracavity power is
sufficiently large: the Rabi angular frequency $\Omega_{\rm Rabi}$
has to be greater than the relaxation timescales of the gain medium
(more precisely, $\Omega_{\rm Rabi}$ is sufficiently greater than
$(T_1 T_2)^{-1/2}$, where $T_1$ is the gain relaxation time). The
instability threshold is often called the second laser threshold due
to its universal nature.

Pioneering theoretical works stimulated extensive experimental
studies that finally resulted in the observation of the Lorenz-type
chaos in a far-infrared single-mode laser \cite{weiss}. However,
despite almost 40 years of efforts, the experimental demonstration
of the multimode RNGH instability has remained somewhat
controversial \cite{Stroud,Hogenboom,Pessina,Viogt, FuHaken,Roldan}.
%
%

\begin{figure*}[htb]
\includegraphics[width=7.4cm]{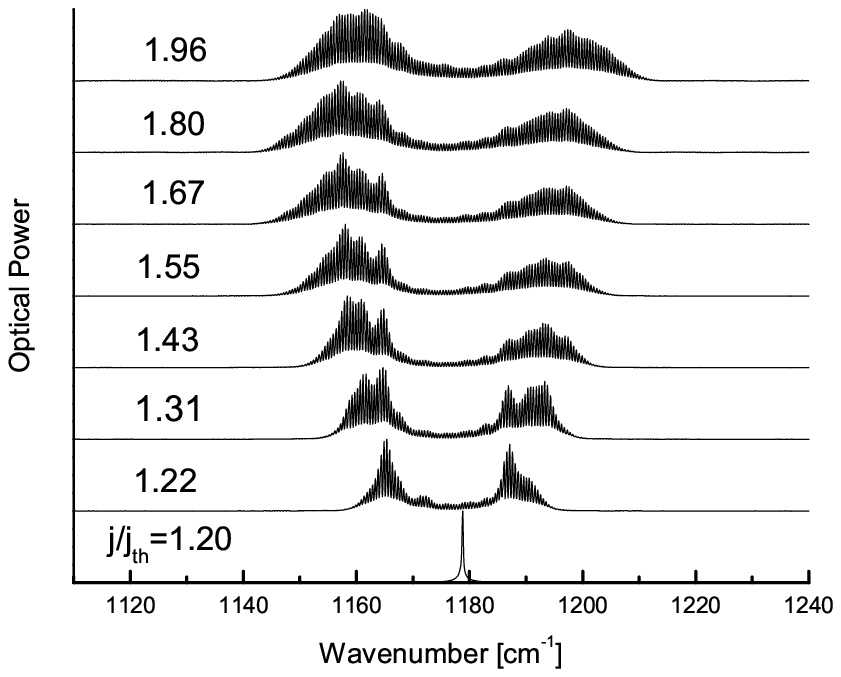}
\includegraphics[width=8cm]{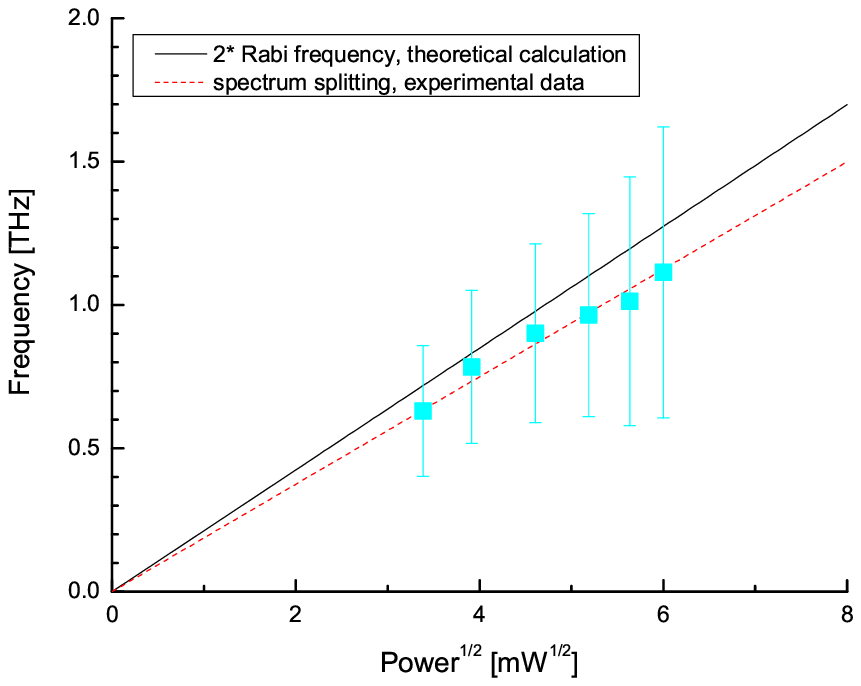}
\caption{(a) Optical spectra vs. pumping ratio ($j/j_{\rm th}$)
above threshold obtained in cw at 300K with a 3 $\mu$m wide buried
heterostructure lasers emitting at 8.38 $\mu$m. For $1<j/j_{\rm
th}<1.2$ the spectra are identical to $j/j_{\rm th}<1.2$. (b)
Spectral splitting and twice the Rabi frequency $\Omega_{\rm
Rabi}/(2\pi)$ vs. square root of output power collected from a
single laser facet. The different quantities reported on the graph
were deduced from the experimental data shown in \ref{fig1} (a). The
dashed line is a least-square linear fit of the data.} \label{fig1}
\end{figure*}

In lasers with long gain recovery compared to the cavity roundtrip
time, the instability caused by a saturable absorber can often lead
to mode locking \cite{HausReview}. When the gain recovery time is
short compared with the cavity round-trip time, it is usually
assumed that laser dynamics becomes very primitive and uninteresting
(so-called class A laser). In this case mode locking is impossible
according to conventional theory, and the relaxation oscillation
frequency becomes purely imaginary \cite{relosc}. Surprisingly, as
we show in this Letter, it is under these conditions that the
elusive RNGH instability can be observed. We show that quantum
cascade lasers (QCLs) are uniquely suited for studying these
coherent effects which, along with spatial hole burning (SHB),
become a key factor in dictating the dynamics of the laser.

QCLs, because they are based on ultrafast tunneling and
phonon-limited intersubband transitions, belong to the class of
lasers which have a extremely fast gain recovery, in the range of a
few picoseconds \cite{CapassoReview}. Recent experiments showed
indeed that the gain recovers within a few picoseconds, which is
approximately an order of magnitude shorter than the cavity
round-trip time \cite{Norris}. Since its invention in 1994, QCLs
have undergone tremendous improvement \cite{CapassoIEEE}. Recent
development of low loss, high power QCLs \cite{Diehl, Diehl2}
enables the study of previously under-investigated aspects, such as
the richness of the optical spectrum and the ultrafast dynamics of
these devices. In Ref.~\cite{Paiella}, strong evidence of
self-pulsations at the cavity roundtrip frequency was reported in
QCLs, in particular a large broadening of the spectrum above the
threshold of this instability was observed. However, no detailed
pulse characterization was provided. The technological potential of
QCLs calls for a better understanding of the interplay of various
instabilities in the parameter regime dictated by these lasers.
Moreover, the Rabi frequency in QCLs at the power levels of a few
hundred milliwatts is of the order of a few THz, much larger than
the spacing of Fabry-Perot modes. Therefore coherent effects should
be easily observable in QCLs.

In this Letter we present a clear experimental demonstration of a
coherent instability, driven by the same mechanism as the RNGH
instability. It is identified in the most direct manner, by
demonstrating in the optical spectrum of QCLs a splitting
corresponding to twice the Rabi frequency. To the best of our
knowledge, this is the first observation of the RNGH mechanism in a
semiconductor laser and even more generally in a solid-state laser.

The instability observed differs in some respects from the original
RNGH instability \cite{Risken, Haken}. The threshold of instability
is as low as $\sim$ 50\% above the laser threshold. In addition, the
pure RNGH instability typically gives rise to spectra with one
central mode and two sidebands separated from it by the Rabi
frequency, whereas in our experiments we observed two peaks only,
similarly to Ref.~\cite{Stroud}. However the mechanism of the
instability is the same in essence, namely the Rabi oscillations of
the population inversion due to coherent effects. The differences
from the RNGH instability as it occurs in ideal conditions
\cite{Risken,Haken} can be attributed to the presence of saturable
absorption and SHB.

The QCLs studied were fabricated from two different wafers (wafer \#
3251 and 3252) grown by metalorganic vapor phase epitaxy. The
devices were processed into buried heterostructure lasers, in which
an insulating Fe-doped InP layer is regrown
after etching of the ridges \cite{Diehl,Diehl2}. 
The active region of all the samples tested is based on a
four-quantum-well design, which rely on a double phonon resonance to
achieve population inversion \cite{Beck}. Note however that the
multimode operation described in the present letter was also
observed with lasers based on so-called three-quantum-well designs
\cite{CapassoReview}. Fig. \ref{fig1}(a) shows the optical spectra
of a laser operated in continuous wave (cw) at room temperature. The
active region of this laser is 3$\mu$m wide and its emission
wavelength is close to 8.38$\mu$m (wafer \#3251). The laser was
cleaved into a 3.25mm long bar and soldered with Indium onto a
copper heat sink. The spectra were measured by a Fourier transform
infrared spectrometer (FTIR) equipped with a deuterated triglycine
sulphate (DTGS) detector.

\begin{figure}[htb]
\includegraphics[width=8cm]{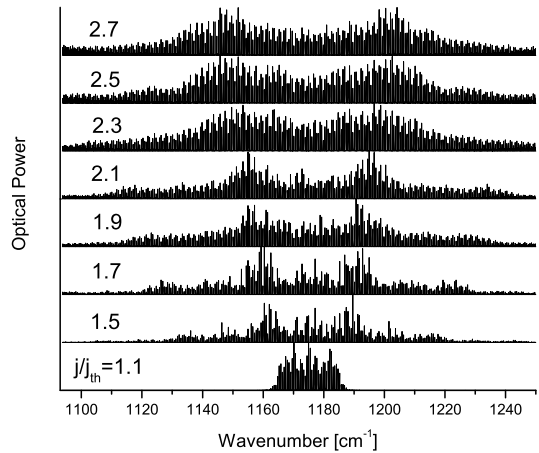}
\caption{Results of numerical simulations of the spectra based on
the Maxwell-Bloch equations including a saturable absorber and
spatial hole burning for different values of the current density
normalized to the threshold value.}\label{fig3}
\end{figure}

As shown in Fig.~\ref{fig1} (a), the laser spectrum is single mode
close to threshold and broadens as the pumping current increases,
splitting into two separated humps. The difference between the
weighted centers of the two peaks increases linearly as a function
of the square root of the collected output power from one facet, as
shown in Fig.~\ref{fig1} (b) (square dots with the dashed line as
its best-fit). The Rabi angular frequency $\Omega_{\rm Rabi}$ can be
easily calculated using the formula $\Omega_{\rm Rabi}=\mu E/\hbar
=\mu \sqrt{2nI_{\rm ave}/(c\epsilon)}/\hbar$, where $\mu$ is the
electron charge times the matrix element of the laser transition
(=2.54nm). $I_{\rm ave}$ is the average intracavity intensity in the
gain region, which can be derived from the measured output power.
$c$ is the speed of light in vacuum and $n$ is the background
refractive index. For all the values of the intensity corresponding
to the spectra reported in Fig.~\ref{fig1}(a), $\Omega_{\rm Rabi}$
was calculated, multiplied by a factor two and then added to
Fig.~\ref{fig3} (b) (solid line). A very good agreement is found
between the experimental splitting and twice the estimated Rabi
frequency. Both curves fall indeed well within the error bars
\cite{errorbars}. As mentioned before, the theory behind the RNGH
instability predicts that the large intracavity intensity will
result in parametric gain at frequencies detuned from the maximum of
the gain curve by the Rabi frequency. The agreement mentioned above
is thus a strong indication of the RNGH instability in QCLs.

In order to better understand the experimental spectra of the QCLs
presented in Fig.~\ref{fig1} (a), we use a simple model based on the
standard one-dimensional Maxwell-Bloch equations \cite{Eberly},
where the active medium is described by an ``open" two level system
\cite{Boyd}. However contrary to the standard unidirectional
Maxwell-Bloch equations, we allow the electromagnetic field to
propagate in both directions. The waves traveling in the two
directions are coupled, as they share the same gain medium. This
gives rise to SHB \cite{Yariv}: The standing wave formed by a cavity
mode imprints a grating in the gain medium through gain saturation.
As a result, other modes become more favorable for lasing, and a
multimode operation is triggered.

In the slowly varying envelope approximation, the equations read:
\begin{subequations}\label{allreqs}
\begin{eqnarray}\label{MB1}
\frac n c
\partial _tE_\pm&=&\mp \partial_zE_\pm -\ii \frac{k N \mu \Gamma}{2\epsilon_0 n^2} \eta_{\pm}-\frac 1 2
\ell(E_+,E_-)E_\pm\phantom{AA}
\\ \partial _t \eta_\pm&=&\frac{\ii \mu}{2\hbar}(\Delta_0 E_\pm+
\Delta_2^\pm E_\mp) -\frac{\eta_\pm}{T_2}\\ \partial _t
\Delta_0&=&\frac{\Delta_p-\Delta_0}{T_1}+\frac{\ii \mu}{\hbar}(E
^*_+ \eta_+ +E ^*_- \eta_- - c.c.)  \\ \label{4} \partial
_t\Delta_2^\pm&=&\pm\ii\frac \mu \hbar(E ^*_+ \eta_-- \eta^*_+ E_- )
-\frac{\Delta_2^\pm}{T_1}.
\end{eqnarray}
\end{subequations}
The + and $-$ subscripts label the two directions of propagation.
$E$ and $\eta$ are the slowly-varying envelopes of the electric
field and the polarization respectively. The actual electric field
and polarization are obtained by multiplying $E$ and $\eta$ by
$e^{i\omega t}$ ($\omega$ is the optical resonance frequency) and
taking the real part. The position-dependent inversion is written as
the sum of the three terms, $\Delta_0$, $\Delta_2^+e^{2ikz}$, and
$\Delta_2^-e^{-2ikz}$, where $(\Delta_2^+)^*\equiv\Delta_2^-$. The
inversion is thereby represented by two slowly varying functions
($\Delta_0$ and $\Delta_2^+$), and $e^{\pm2ikz}$ gives the fast
variation in space. All the quantities mentioned so far are
functions of space $z$ and time $t$.

$\ell(E_+,E_-)$ is the loss in the cavity (not including the mirror
loss), which is allowed to be nonlinear and to dependent on the
intensity. In this work we assume
\begin{equation}\label{loss}
 \ell(E_+,E_-)=\ell_0-\gamma(|E_+|^2+|E_-|^2),
\end{equation}
where $\ell_0$ is the linear loss and $\gamma$ is the self-amplitude
modulation coefficient characterizing the nonlinear (saturable) part
of the loss. Such a saturable absorbtion mechanism can come from
Kerr-lensing \cite{HausReview,Paiella}, caused by a nonlinear
refractive index $n_2$ in the active region. As the intensity
increases, the mode is more confined in the plane transverse to the
propagation direction, and the net gain it experiences is greater.
The reason is twofold: First, the mode overlaps more with the active
region, leading to a larger modal gain (this mechanism is often
called ``soft Kerr-lensing" \cite{Piche}). Second, the overlap with
the metal contacts is smaller, leading to smaller losses
\cite{Paiella}.

$E_+$ and $E_-$ satisfy the boundary conditions $E_+=rE_-$ at the
$z=0$ boundary and $rE_+=E_-$ at the $z=L$ boundary ($L$ is the
cavity length and $r\approx 0.53$ is the reflection coefficient).
The other quantities in \eq{allreqs} are constants: $k$, $N$, and
$\Gamma$ are the wavenumber (in the material) associated with the
resonance optical frequency, the electron density in the active
region, and the overlap factor between the laser mode and the active
region respectively.

Figure \ref{fig3} shows spectra that were obtained by solving
numerically the equations \eq{allreqs} with the following
parameters: for the saturable absorber, we used $\gamma=
10^{-8}\rm\frac{cm}{V^2}$, obtained from two-dimensional mode
 simulations, assuming a $n_2\approx 10^{-9}\rm
\frac{cm^2}{W}$ \cite{Paiella}. The index change due to this $n_2$
at typical intracavity intensities is about $10^{-3}$. The other
parameters are $\ell_0=5\rm\, cm^{-1}$, $T_1=0.5$ ps \cite{Norris},
$T_2=0.07$ ps (corresponding to a gain FWHM bandwidth of 4.8 THz),
$L=0.3$ cm, and $n=3$, which are typical values for these lasers.
$N$ and $\Gamma$ are not needed as long as the pumping is expressed
relative to the lasing threshold. Note that the simulated spectra
presented in Fig. \ref{fig3} are averaged over about a microsecond.
Only then does the average spectrum reach a steady state and a clear
pattern shows up. The averaging is motivated by the fact that
experimentally the spectra are acquired over an even much longer
timescale. The envelopes of the spectra show two clear peaks whose
separation compares well with twice the Rabi frequency, similarly to
the experiment.

\begin{figure}[htb]
\includegraphics[width=8cm]{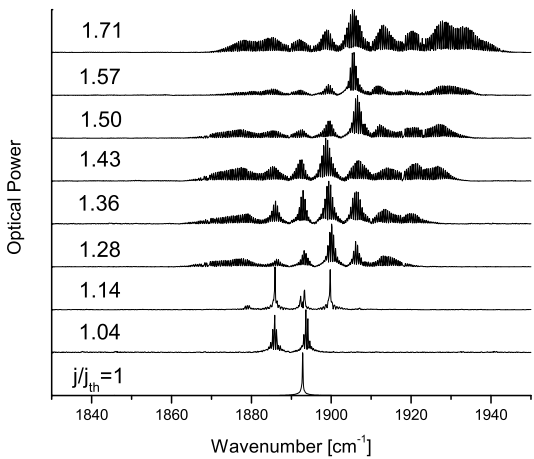}
\caption{Optical spectra vs. pumping ratio above threshold obtained
in cw at 300K with a 5 $\mu$m wide buried heterostructure lasers
emitting at 5 $\mu$m.}\label{fig2}
\end{figure}

The lowering of the RNGH instability threshold by a saturable
absorber can be established analytically by means of linear
stability analysis. We propose this mechanism as the main reason for
the observation of the RNGH instability at much lower pumping than
RNGH theory predicts. In order to support this idea, we now present
spectra from another device similar to the one described previously.
The only difference  between the two lasers is a shorter optical
wavelength (5$\mu$m) (wafer \#3252) and a wider active region
(5$\mu$m). The two-dimensional waveguide simulations indicate a much
weaker Kerr-lensing effect in these QCLs ($\gamma$ is smaller by a
factor of 4), due to the much larger ratio of active region width to
wavelength. The measured optical spectra obtained at 300K in cw mode
with the 5$\mu$m device are shown in Fig.~\ref{fig2}. The data
clearly show that the laser is at first single mode close to
threshold and becomes multimode immediately after a slight increase
of the pumping current. The envelopes of the spectra consist of
multiple peaks, with an average separation 0.2THz, independently of
the pumping. Numerical integration of \eq{allreqs} without a
saturable absorber ($\gamma=0$) leads to spectra that qualitatively
agree with the ones in Fig.~\ref{fig2}.

Ref.~\cite{FuHaken} suggested that the suppression of the central
peak in RNGH-type spectra can be due to the complex level structure
of the gain medium, a dye molecule in that case. We show that SHB
can also result in the suppression of the central peak
(Fig.~\ref{fig3}).

Our postulation of saturable absorption due to Kerr-lensing is
supported by more extensive study of different devices beyond those
shown in this Letter. First, we observed that for the same emission
wavelength, a broad active region leads to a less pronounced
RNGH-type signature. Second, we have also tested several standard
ridge waveguide QCLs, for which the sidewalls of the ridges are
covered by the gold contact. For these devices the coupling between
the optical mode and the metal is expected to be stronger and so is
the effect of saturable absorber due to Kerr-lensing. The spectral
behavior observed in this class of devices is dominated by RNGH-type
instability.

In summary, a coherent multimode instability in quantum cascade
lasers (QCLs) has been observed. It is similar in many ways to the
Risken-Nummedal-Graham-Haken (RNGH) instability. The threshold of
the observed phenomenon is significantly lower than in the original
RNGH instability, which is attributed to the presence of a saturable
absorber in the laser. For devices with a weaker saturable absorber,
the envelope of the optical spectrum consists of many maxima whose
separations are independent of the intracavity power. The nontrivial
shape of the spectrum can be explained by SHB.

Support from the U.S. Army Research Laboratory and the U.S. Army
Research Office under grant number W911NF-04-1-0253 is gratefully
aknowledged. Part of the device processing was done at the Center
for Nanoscale Systems (CNS) at Harvard University. Harvard-CNS is a
member of the National Nanotechnology Infrastructure Network (NNIN)


\begin{thebibliography}{10}
\bibitem{errorbars} The main contribution to the error bars is due
to the width of the two peaks in the specra.
\end{thebibliography}
\end{document}